\documentclass[a4paper]{article}
\usepackage{amsfonts}
\usepackage{amsmath,graphicx, mathtools}
\usepackage{amssymb}
\usepackage{array}
\usepackage{INTERSPEECH2022}
\usepackage{enumerate}
\usepackage{enumitem}

\newcommand{\frontend}{\mathcal{F}}

\title{Learning neural audio features without supervision}
\name{Sarthak Yadav$^1$, Neil Zeghidour$^2$}
\address{
  $^1$University of Glasgow, UK\\
  $^2$Google Research, Paris, France}
\email{s.yadav.2@student.gla.ac.uk, neilz@google.com}

\begin{document}

\maketitle
\begin{abstract}
Deep audio classification, traditionally cast as training a deep neural network on top of mel-filterbanks in a supervised fashion, has recently benefited from two independent lines of work. The first one explores ``learnable frontends'', i.e., neural modules that produce a learnable time-frequency representation, to overcome limitations of fixed features. The second one uses self-supervised learning to leverage unprecedented scales of pre-training data. In this work, we study the feasibility of combining both approaches, i.e., pre-training learnable frontend jointly with the main architecture for downstream classification. First, we show that pretraining two previously proposed frontends (SincNet and LEAF) on Audioset drastically improves linear-probe performance over fixed mel-filterbanks, suggesting that learnable time-frequency representations can benefit self-supervised pre-training even more than supervised training. Surprisingly, randomly initialized learnable filterbanks outperform mel-scaled initialization in the self-supervised setting, a counter-intuitive result that questions the appropriateness of strong priors when designing learnable filters. Through exploratory analysis of the learned frontend components, we uncover crucial differences in properties of these frontends when used in a supervised and self-supervised setting, especially the affinity of self-supervised filters to diverge significantly from the mel-scale to model a broader range of frequencies.
\end{abstract}

\section{Introduction}
\label{sec:intro}
Mel-filterbanks have long remained the features of choice for machine learning and audio signal processing. Mel-filterbanks first pass a spectrogram through a bank of triangular bandpass filters that are logarithmically scaled to model human perception of pitch \cite{stevens1940relation}, referred to as the mel-scale, providing shift-invariance and robustness to deformations \cite{anden2014deep}. The dynamic range of the resulting coefficients is then typically compressed by a logarithm, to replicate human sensitivity to loudness. However, mel-filterbanks suffer from inherent limitations of fixed features. Not only the mel-scale has been reconsidered several times over its history \cite{o1987speech, umesh1999fitting} but logarithmic compression has also been shown to be suboptimal with respect to $n^{th}$ root non-linearities \cite{schluter07, lyons08}. These limitations, paired with the development of deep learning methods, have fostered a growing corpus of work on learning an audio frontend from raw waveform signals. Most of these contributions focus on learning a filterbank as an alternative to mel-scale triangular filters \cite{palaz2015convolutional, sainath2015learning, balestriero2018spline, zeghidour2018end, zeghidour2018learning, ravanelli2018speaker, noe2020cgcnn}, while some propose replacing logarithmic compression by a trainable non-linearity \cite{pcen, pcen_why_how}, or learning all operations (filtering, pooling, compression) in an end-to-end fashion \cite{zeghidour2021leaf}. 

Learnable alternative to fixed features are typically trained and evaluated in a supervised setting. In that context, the lack of strong priors (e.g., mel-scale, log compression) is compensated by strong labelling of the training data, which allows exploring the space of models to reach a parameterization that can consistently outperform fixed features \cite{zeghidour2021leaf}. In recent years, there has been a rise in learning audio representations with self-supervision \cite{oord2018representation, baevski2020wav2vec, tagliasacchi2020pre, hsu2021hubert,  niizumi2021byol, wang2021towards, saeed2021contrastive}, capable of learning general-purpose acoustic representations without extensive manual data annotation and have demonstrated excellent few-shot learning performance. However, these methods either adopt large-scale architectures without any structure that would be akin to audio features \cite{oord2018representation, baevski2020wav2vec} or rely on mel-filterbanks \cite{tagliasacchi2020pre, fonseca2021unsupervised, saeed2021contrastive}. Thus, the question of learning strong audio frontends without labelled data still remains unanswered.

In this work, we explore learning audio features in a self-supervised setting, to study whether 1) learnable audio frontends can outperform audio features even without labelled data 2) at convergence, the learned filterbanks and compression non-linearity differ between the supervised and the self-supervised setting. More specifically, we explore two previously proposed learnable audio frontends: SincNet \cite{ravanelli2018speaker}, and LEAF \cite{zeghidour2021leaf}, both in a supervised multi-label classification setting, and in a self-supervised framework using contrastive learning, on Audioset \cite{gemmeke2017audio}. On a linear probe transfer task, learnable frontends significantly outperform fixed mel-filterbanks, confirming experimental results previously limited to the supervised setting \cite{ravanelli2018speaker, zeghidour2021leaf}. When exploring the role of filter initialization, we observe that while randomly initialized LEAF and SincNet frontends do not improve supervised learning performance, they surprisingly improve performance for self-supervised training over their mel-scale initialized counterpart, a counter-intuitive result that suggests reconsidering the mel-scale. Moreover, frequency analysis of the learned filters shows that they converge to similar configurations in a supervised setting, regardless of their initialization. In contrast, self-supervised learning converges to more diverse configurations, with the randomly initialized kernels modelling a broader range of frequencies. In the upcoming sections, we present the proposed approach, followed by an in-depth analysis of our experimental setting and results on the benchmarked datasets. Finally, we conclude with an inspection of the learned components of the audio frontends.


\section{Method}
\label{sec:method}

The proposed model has two core components: i) a common \textit{neural backbone} which consists of an audio frontend followed by an encoder, and ii) a task-dependent shallow MLP head. Our model is trained either in a supervised or a self-supervised setting, and is described further in this section.

\subsection{Audio frontends}
\label{ssec:frontends}

\begin{figure}
  \centering
  \includegraphics[width=0.8\linewidth]{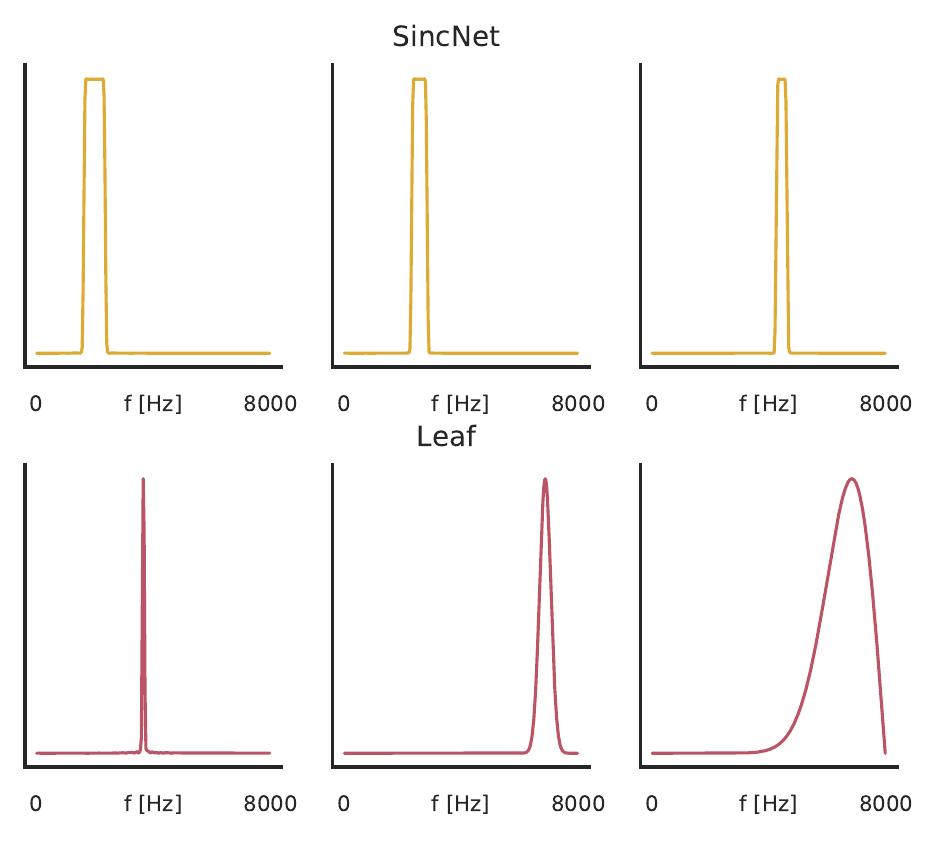}
  \caption{Frequency response of randomly selected SincNet filters and LEAF filters, with truncated normal and uniform initialization schemes, respectively.}
  \label{fig:frontends}
\end{figure}

A frontend $\frontend_\psi$ maps an input waveform $x \in \mathbb{R}^T$ sampled at a frequency $F_s$ (Hz) to a $({M \times N})$-dimensional feature space, where $M$ denotes the number of temporal frames and $N$ represents the number of frequency bins. Along with log-compressed mel-filterbanks as our baseline, we explore two recent learnable audio frontends, SincNet and LEAF. 

\noindent\textbf{SincNet} \cite{ravanelli2018speaker} utilizes learnable sinc function based band-pass filters with a rectangular frequency response (see Figure~\ref{fig:frontends}) that are parameterized by cut-off frequencies $[f_1, f_2]$:
\begin{equation}
    g[n, f_{1}, f_{2}] = 2f_{2}sinc(2\pi f_{2}n) - 2f_{1}sinc(2\pi f_{1}n),
\end{equation}
where $sinc(x) = sin(x)/x$, and the cut-off frequencies are initialized on the mel-scale. This is followed by a leaky ReLU non-linearity and a max-pooling layer for down-sampling.

\noindent\textbf{LEAF} \cite{zeghidour2021leaf} proposes a learnable frontend that utilizes convolution with $N$ learnable complex-valued Gabor filters with a gaussian frequency response (Fig~\ref{fig:frontends}), each of length $W$, which are parameterized by their center frequencies $\eta_n$ and inverse bandwidths $\sigma_n$ as follows:
\begin{equation}
    \varphi_n(t) = e^{i 2\pi \eta_n t} \frac{1}{\sqrt{2\pi}\sigma_n} e^{-\frac{t^2}{2\sigma_n^2}},
\end{equation}
where $n = 1, \ldots ,N$ and $t = -W/2, \ldots, W/2$, and the center frequencies are initialized in the $[0, 1/2]$ range (normalized units) on the mel-scale.
This is followed by a squared modulus operator, which gets the input back to the real-valued domain, followed by a learnable Gaussian low-pass pooling layer. Finally, a Per-Channel Energy Normalization layer \cite{pcen}, applies learned compression and normalization to the coefficients. Hence, while SincNet only learns the filterbank, LEAF jointly learns filtering, pooling, compression and normalization. 

\noindent\textbf{Random initialization schemes}: Previous work \cite{sainath2015learning, zeghidour2018end} has examined how random initialization affects performance for learnable audio frontends, albeit for other frontends and only in the supervised setting. In the same light, we explore using filters randomly initialized from uniform and truncated-normal distributions for both the SincNet and LEAF audio frontends across both the supervised and the self-supervised training regimes.

\subsection{Backbone architecture}
\label{ssec:arch}

The complete \textit{neural backbone} $f(.)$ consists of the audio frontend $\frontend_\psi$ followed by an encoder $enc(.)$, and maps the input waveform $x \in \mathbb{R}^T$ into a latent representation $h = f(x) = enc(\frontend_\psi(x)) \in \mathbb{R}^D$ and is common across all the experiments. A linear classifier head is added on top of the neural backbone in the supervised setting. We use the lightweight EfficientNetB0 \cite{tan2019efficientnet} network with around $4M$ parameters as the convolutional encoder in this work.

\subsection{Self-supervised pre-training}
\label{ssec:sslpre}

For self-supervised pretraining, we use COLA \cite{saeed2021contrastive}, a contrastive approach for learning general-purpose audio representations. Given the similarity function 
\begin{equation}
    s(x,x') = g(f(x))^\top ~W~ g(f(x')),
\end{equation}
where $f(.)$ is the \textit{neural backbone}, $g(.)$ is a shallow neural network that maps $h$ onto a space $z = g(h) \in \mathbb{R}^G$, and $W\in\mathbb{R}^G$ are bilinear similarity parameters, COLA learns a latent space such that the similarity $s(x, x^+)$ between an anchor-positive pair is greater than the similarity $s(x, x^-)$ between the same anchor and negative samples (unrelated examples) by optimizing the following objective function:
\begin{equation}
    \mathcal{L} = -\log \frac{\exp\left( {\rm s}(x, x^+)\right)}
   {\sum\limits_{\mathclap{x^- \in {\cal X^-}(x)\cup\{x^+\}}} \exp\left( {\rm s}(x, x^-)\right) },
\end{equation}
where ${\cal X^-}(x)$ refers to the set of negative samples. As opposed to existing methods that use perturbations on the anchor/positive samples \cite{chen2020a, fonseca2021unsupervised, wang2021towards}, in COLA, the anchor and its corresponding positive samples are segments from the same audio clip without any perturbations applied, while segments from all other clips in the batch act as negatives, removing the need for benchmarking multiple augmentation strategies.

\section{Experiments}
\label{sec:experiments}


To evaluate the relative improvements of learnable frontends over fixed features based on the level of supervision, we compare SincNet, LEAF and mel-filterbanks in two settings: 1) Supervised training on the AudioSet dataset \cite{gemmeke2017audio}, and 2) self-supervised COLA pre-training on AudioSet followed by linear-probing (only training a linear classifier on top of a frozen backbone) on the SpeechCommandsV2 dataset \cite{warden2018speech}. We choose SpeechCommandsV2 given that it has a moderate number of fixed-sized data samples (over 105000, 1-second audio clips across 35 classes) which simplifies experimentation, and as shown previously in \cite{saeed2021contrastive} it poses a challenging transfer task.

\subsection{Implementation details}
\label{ssec:implementation}

The input signal sampled at $F_s = 16$ kHz is used for all experiments. Log-compressed mel-filterbanks with 64 filters with a window of 25 ms and a stride of 10 ms are used as a baseline. To facilitate comparison, LEAF and SincNet frontends also have $N=64$ learnable filters with kernels of width $W = 401$ ($\approx$ 25 ms at 16 kHz), and their corresponding pooling layers are initialized with a stride of 160 samples (10 ms at 16 kHz). For per-channel energy normalization, the ``sPCEN" variant as proposed in \cite{zeghidour2021leaf} is used, and is here on referred to simply as ``PCEN". To study how the per-channel energy normalization in the LEAF frontend affects performance and model characteristics, we also run an extra set of experiments with a fixed PCEN layer for each initialization scheme.

The neural \textit{backbone} $f(.)$ is common for all the experiments, maps into an embedding of size $1280$ and uses stochastic depth \cite{huang2016deep} for regularization. All supervised AudioSet models are trained on $5$-second random crops with a batch size of 1024 for a total of 50 epochs, with SpecAugment \cite{park2019specaugment} and MixUp \cite{zhang2017mixup}. For COLA pre-training, randomly cropped 960 ms segments from an input clip are used as the anchor-positive pairs and passed through the backbone $f(.)$. We use a projection head $g(.)$ with $G = 512$ units, followed by layer normalization \cite{ba2016layer} and tanh activation. All COLA models are trained with a batch size of 2048 to provide a wider set of negative samples, for 50 epochs. For linear-probe evaluation on the SpeechCommandsV2 dataset, a linear classifier is trained directly on top of the COLA pre-trained neural backbone.

We train all models with an AdamW optimizer \cite{loshchilov2018decoupled}, with a linear warmup to a base learning rate of 5e-4 followed by a cosine decay \cite{loshchilov2016sgdr} to $0$ on a single TPU-v3 machine and repeated at least thrice. 

\subsection{Results}
\label{ssec:expresults}

\begin{table}[t]
  \caption{Test mAP, mAUC and d-prime ($\pm$ std over 3 runs) for supervised classification on AudioSet. LEAF* denotes LEAF frontend with a fixed PCEN layer.}
  \label{tab:audioset}
  \centering
  \setlength{\tabcolsep}{5pt}
  \footnotesize
  \begin{tabular}{lcccc}
    \toprule
    frontend & init & mAP & mAUC & d'\\
    \midrule
    mfbanks & N/A & \textbf{0.380±0.001} & \textbf{0.971±0.001} & \textbf{2.660±0.010}\\
    \midrule
    SincNet & mel & 0.358±0.008 & 0.967±0.001 & 2.593±0.022\\
    SincNet & uniform & 0.351±0.007 & 0.965±0.000 & 2.567±0.006\\
    SincNet & truncn & 0.353±0.007 & 0.966±0.001 & 2.578±0.019\\
    \midrule
    LEAF* & mel & 0.373±0.002 & 0.969±0.001 & 2.639±0.019\\
    LEAF* & uniform & 0.358±0.003 & 0.967±0.001 & 2.595±0.019\\
    LEAF* & truncn & 0.358±0.008 & 0.967±0.001 & 2.593±0.022\\
    \midrule
    LEAF & mel & \textbf{0.380±0.002} & \textbf{0.970±0.001} & \textbf{2.653±0.019}\\
    LEAF & uniform & 0.365±0.003 & 0.968±0.001 & 2.616±0.016\\
    LEAF & truncn & 0.368±0.002 & 0.968±0.000 & 2.622±0.011\\
    \bottomrule
  \end{tabular}
\end{table}

\begin{table}[t]
  \caption{ Test accuracy ($\pm$ std over 3 runs) of COLA pre-training + linear-probe results on SpeechCommandsV2. LEAF* denotes a LEAF frontend with a fixed PCEN layer.}
  \label{tab:linearprobe}
  \centering
  \setlength{\tabcolsep}{5pt}
  \footnotesize
  \begin{tabular}{lcccc}
    \toprule
    frontend & init & Accuracy\\
    \midrule
    mfbanks & N/A & 61.2±2.6\\
    \midrule
    SincNet & mel & 61.6±2.8\\
    SincNet & uniform & 64.5±2.2\\
    SincNet & truncn & 64.2±3.6\\
    \midrule
    LEAF* & mel & 57.4±3.2\\
    LEAF* & uniform & 62.5±2.3\\
    LEAF* & truncn & 63.6±2.8\\
    \midrule
    LEAF & mel & 75.0±1.2\\
    LEAF & uniform & \textbf{76.1±3.3}\\
    LEAF & truncn & 75.8±0.4\\
    \bottomrule
  \end{tabular}
\end{table}

Table~\ref{tab:audioset} reports results on the supervised multi-label classification task on Audioset for all the frontends. The LEAF frontend (row 2) matches the performance of log-compressed mel-filterbanks. For both the learnable frontends, random initialization schemes, although not as good as the default mel-scaled initialization, work remarkably well. It is worth noting that LEAF outperforms SincNet across the board, with even the random initialization schemes performing better than the default mel-scaled SincNet frontend. Overall, our observations on using strong priors (mel-scale) for initialization confirms the findings of previous work \cite{sainath2015learning, ravanelli2018speaker, zeghidour2018learning}.

\begin{figure}[t]
  \centering
  \includegraphics[width=\linewidth]{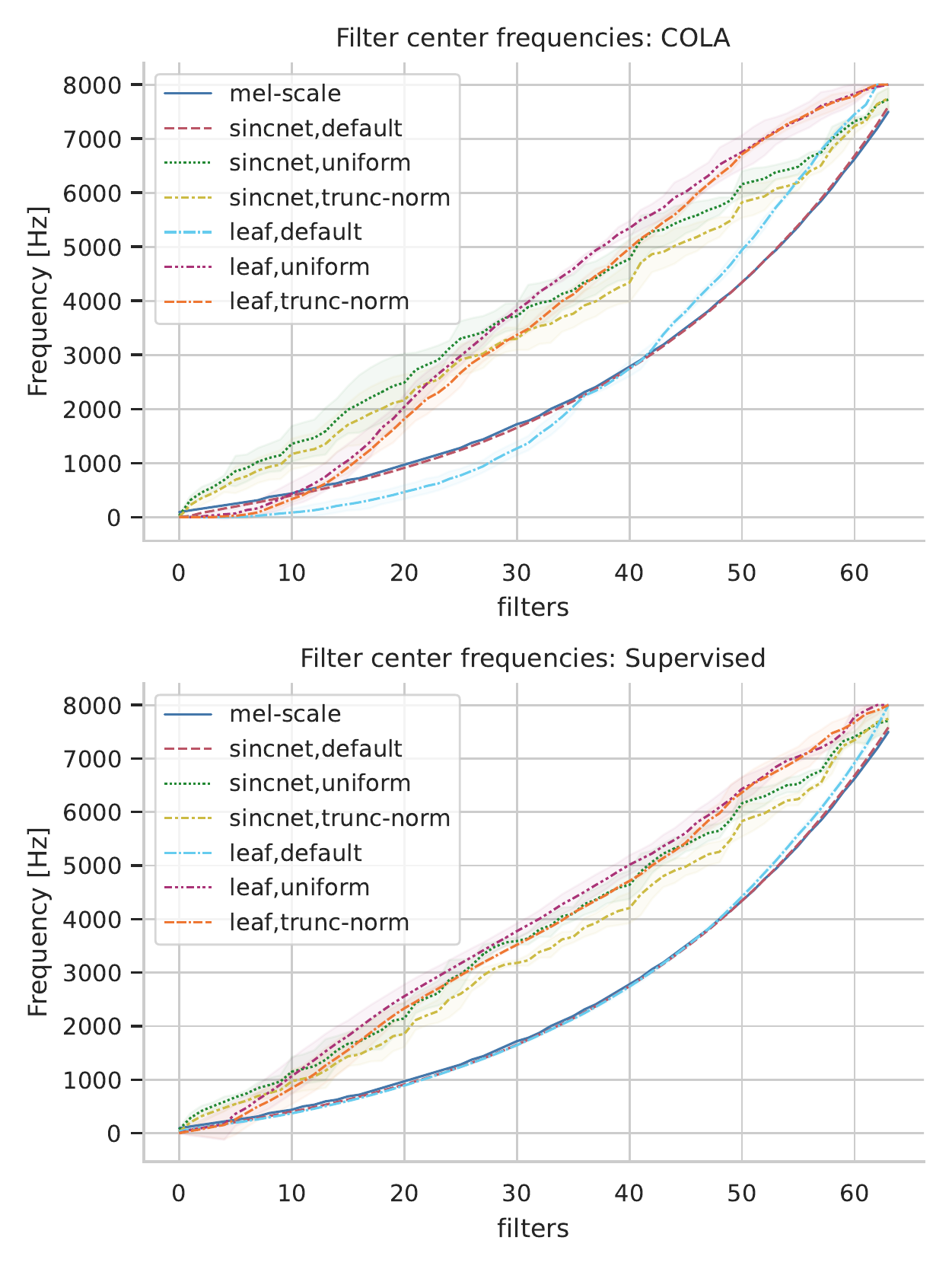}
  \caption{Learned filter center frequencies for LEAF and SincNet frontends, for COLA(top) and Supervised (bottom) settings. \textit{Mel} center frequencies added for reference. \textit{default} denotes default mel-scale based initialization. Error bands show standard deviations across three runs (best viewed in electronic format)}
  \label{fig:centerfreqs}
\end{figure}

Table~\ref{tab:linearprobe} reports linear-probe results for speech command classification on pre-trained COLA embeddings. We observe:

\noindent\emph{- Learnable frontends significantly improve downstream accuracy over fixed features}: The learnable frontends provide significant improvements in linear-probe performance over mel-filterbanks, with the mel-scale initialized LEAF frontend offering a 14\% absolute increment in classification accuracy.

\noindent\emph{- Random initialization outperforms mel-scaled initialization}: Surprisingly, for both LEAF and SincNet, the proposed random initialization schemes outperform initializing filter frequencies on the mel-scale, with a more prominent absolute improvement of $\approx~3\%$ observed for SincNet. Not only is this trend inversely correlated with the supervised results, but it is also counter-intuitive as initializing a filterbank on an auditory scale should provide a stronger starting point for the optimization process than initializing randomly. 

\noindent\emph{- Importance of a trainable PCEN layer}: It is expected that a fixed PCEN (all experiments marked LEAF*) will adversely affect performance, as observed for the supervised setting in Table~\ref{tab:audioset}. However, as evident from Table~\ref{tab:linearprobe}, the adverse effect on performance is even more significant in the self-supervised setting, with a drastic reduction in linear-probe accuracy when a fixed PCEN layer is used. To the best of our knowledge, this work is the first to show the benefits of learning PCEN compression and normalization in a self-supervised setting.

\subsection{Analysis of the learned frontend components}
\label{ssec:analysis}

\noindent\emph{i. Visualizing filter center frequencies}
\vspace{0.25em}

We visualize center frequencies learned by the first convolution layer in Leaf and SincNet frontends for both COLA and supervised settings across all initialization schemes allowing us to analyze and compare frequencies modelled by the frontends under different settings (see Figure~\ref{fig:centerfreqs}). When training for supervised classification, both frontends converge to center frequencies that follow a scale similar to their initialization (e.g., mel or uniform). On the other hand, self-supervised learning yields much more diverse filterbanks at convergence. It is also worth noting that for the LEAF frontend, even for mel-scale initialization, the center frequencies diverge away much further from the mel-scale for COLA in comparison to supervised training. The fact that contrastive loss draws learnable filters much further from their initialization than cross-entropy suggests that learnable frontends could benefit self-supervised learning even more than supervised classification. 

\begin{figure}[t]
  \centering
  \includegraphics[width=\linewidth]{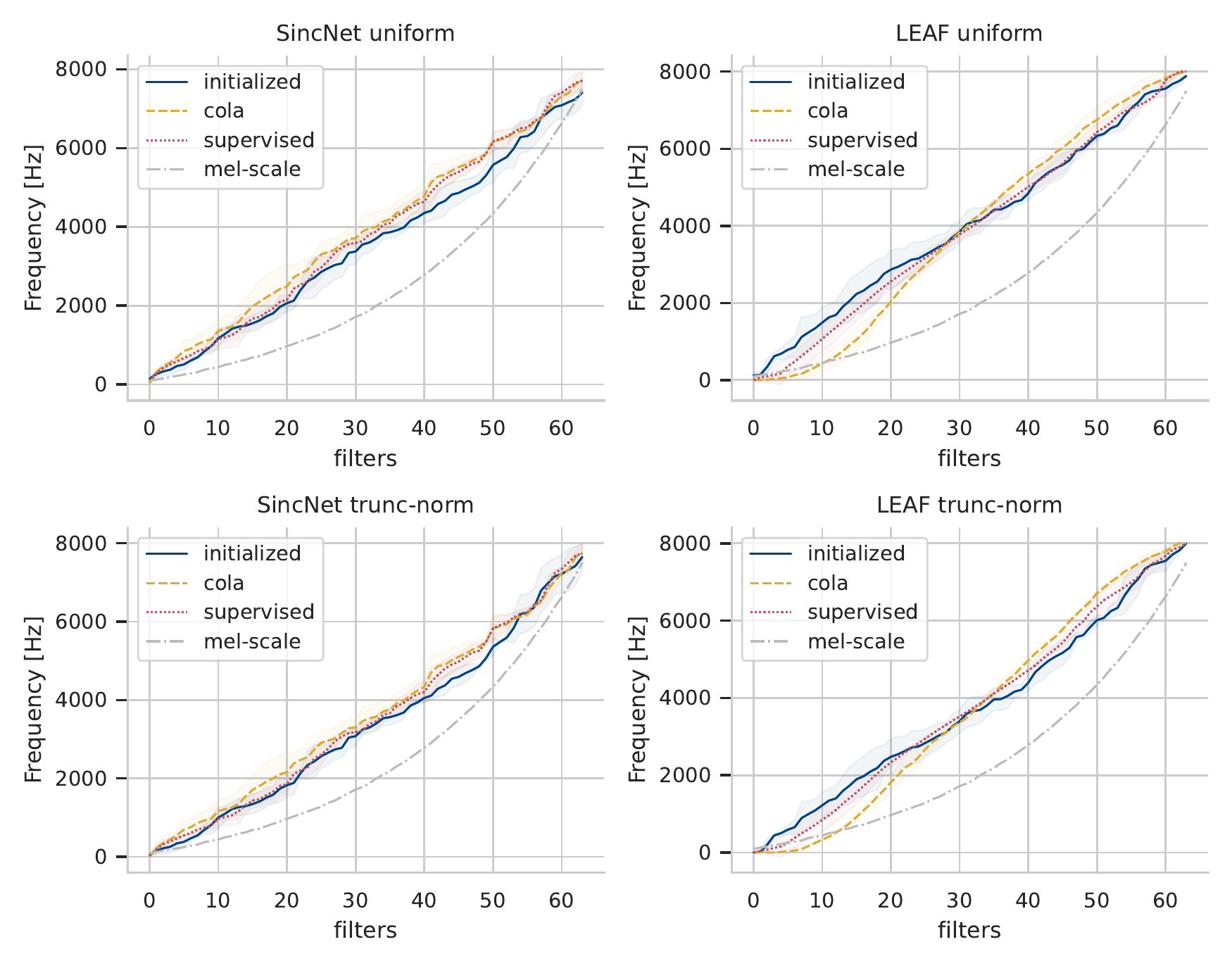}
  \caption{Comparing random initialization schemes at initialization and convergence for the SincNet (left) and LEAF frontends (rights)}
  \label{fig:convergenceleaf}
\end{figure}

\noindent\emph{On random initialization:} Figure~\ref{fig:convergenceleaf} shows randomly initialized LEAF and SincNet frontends, respectively, at initialization and at convergence. SincNet center frequencies at convergence overlap with their initialization throughout the frequency spectrum for both uniform and truncated normal initializations. LEAF filters are much more distinct in their center frequencies modelled across the COLA and supervised training regimes. In the supervised setting, randomly initialized LEAF filters converge closer to their initialization, in contrast to COLA pre-training where they diverge to model a broader set of center frequencies, especially in the 0-2000 Hz and the 3500-7000 Hz range, improving performance over mel-scale initialization.


These observations support empirical observations made in Section~\ref{ssec:expresults}: across frontend configurations, there is a much larger variation in linear-probe performance as compared to supervised training. Moreover, while we could have expected learnable filters to converge to an auditory, logarithmic scale akin to the mel-scale, they instead converge to an almost linear scale which yet provides higher downstream accuracy.

\vspace{0.25em}
\noindent\emph{ii. Self-supervised learning of PCEN}
\vspace{0.25em}

\begin{figure}[t]
  \centering
  \includegraphics[width=\linewidth]{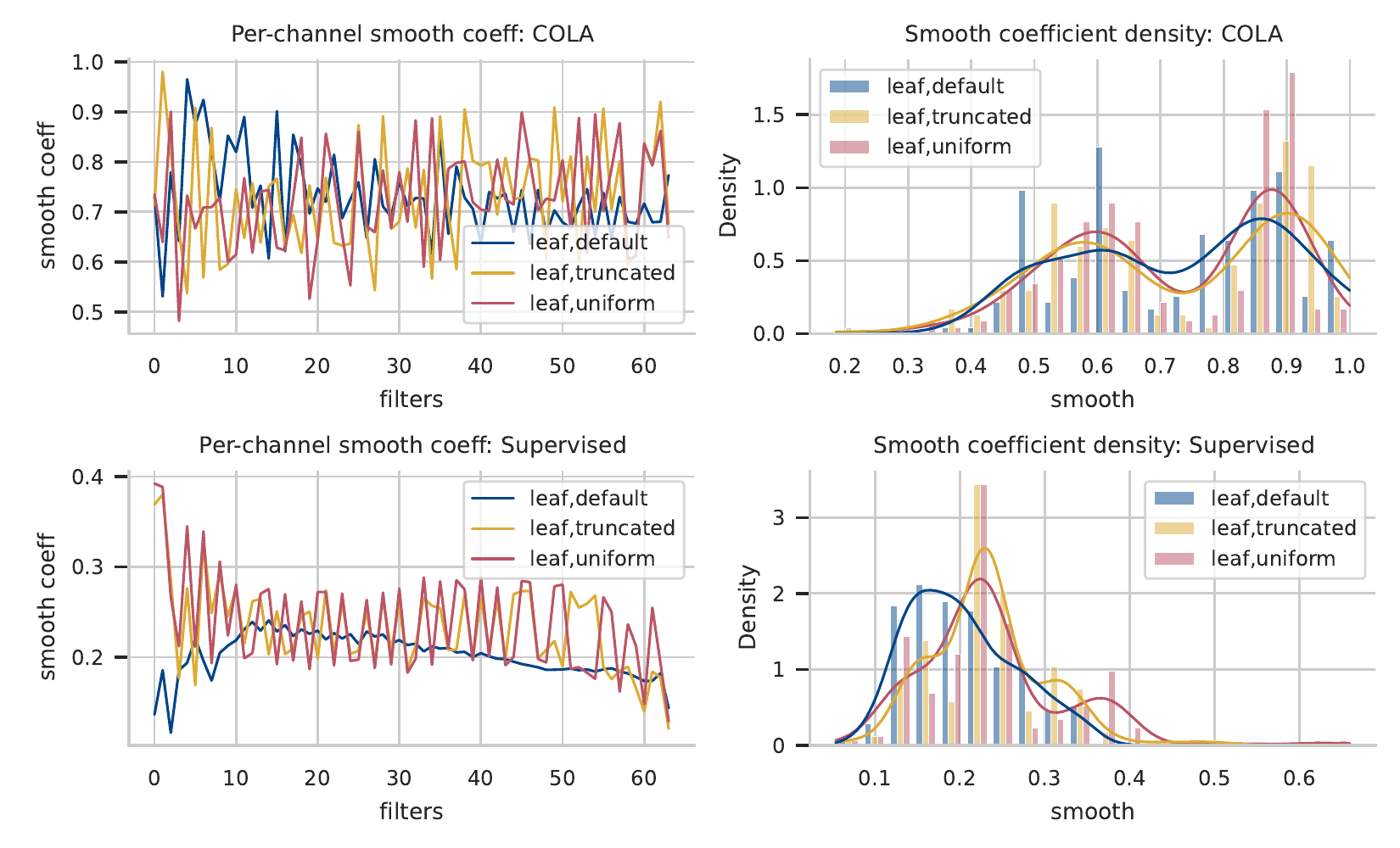}
  \caption{Learned per-channel smoothing coefficients (left) and overall smoothing coefficient density plot (right) of the learned PCEN layer for COLA (top) and Supervised (bottom). Notice the alternating pattern on the left, similar to \cite{pcen}}
  \label{fig:smooth}
\end{figure}

While a PCEN layer learns several parameters including gain normalization and dynamic range compression offset and exponent (often denoted $\alpha$, $\delta$ and $r$, respectively), we focus on the smoothing coefficient $s$, which is the exponential moving average parameter that governs the extent of smoothing of the input time-frequency representation based on its past values. Figure~\ref{fig:smooth} shows per-channel smoothing coefficients as well as their density histogram at convergence. Smoothing parameters tend to oscillate between channels, a pattern similar to that observed by \cite{pcen}, which they posit as an effort by the model to focus on different features to obtain more discriminative information. It's worth noting that this alternating behaviour is more pronounced for the COLA setting and for the randomly initialized frontends. The smoothing coefficient density plot in Figure~\ref{fig:smooth} shows a cleaner picture: COLA learns coefficients that are significantly more spread out over the $[0., 1.]$ range, even for the default mel-scaled initialization, while supervised learning converges to values centered around $0.2$. 

These observations, coupled with the fact that a fixed PCEN layer drastically reduces self-supervised performance (see Sec~\ref{ssec:expresults}), suggest that PCEN possibly enables self-supervised learning to effectively capture discriminative information, and in the examined case, more so than supervised learning.

\section{Conclusion}
In this work, we study the feasibility of training learnable audio frontends in a self-supervised fashion. When pre-training a convolutional encoder jointly with SincNet or LEAF on Audioset, the downstream linear-probe performance on a speech command classification task improves significantly. We conduct an exploratory analysis of the learned filters and normalization coefficients, highlighting major differences in learning dynamics between the supervised and the self-supervised setting. Interestingly, our experiments suggest that while a mel-scale initialization of learnable filters expectedly improves the final performance in the supervised setting, self-supervised learning rather benefits from random initialization. 

\section{Acknowledgements}
This research was supported by the TPU Research Cloud (TRC) program, a Google Research initiative. We also thank Marco Tagliasacchi for helpful suggestions.

\newpage
\bibliographystyle{IEEEtran}
\bibliography{mybib}

\end{document}